# LIQUID-LIKE GROWTH OF COLLOIDAL NANOCRYSTALS BY COALESCENCE

BIN YUAN[1,2†], LUDOVICO CADEMARTIRI[1,2,3,4*]

[1] *Department of Materials Science & Engineering, Iowa State University of Science and Technology, 2220 Hoover Hall, Ames, IA, 50011, USA*

[2] *Department of Chemical & Biological Engineering, Iowa State University of Science and Technology, Sweeney Hall, Ames, IA, 50011, USA*

[3] *Ames Laboratory, U.S. Department of Energy, Ames, IA, 50011, USA*

[4] *Department of Chemistry, Life Sciences, and Sustainability, University of Parma, Parco Area delle Scienze, 17/A, 43124 Parma, Italy*

[†] *Current Address: Mechanical Engineering Department, Carnegie Mellon University, 5000 Forbes Avenue, Pittsburgh, PA, 15213, USA*

[*] *Author to whom correspondence should be addressed: ludovico.cademartiri@unipr.it*

# SUMMARY PARAGRAPH

Our understanding of the growth of crystals is dominated by the classical description according to which individual atoms or molecules, driven by supersaturation, add to crystal facets. As a result, the growth of hard matter is still mostly considered to be fundamentally incomparable to the growth of soft matter, like polymers or liquids.

By a combination of experiment and modeling we here show how amine-capped PbS colloidal nanoparticles grow in the absence of supersaturation by coalescence, like droplets in an emulsion. Specifically, we (i) determine that the rates of crystal-crystal coalescence are remarkably high ($10^{-2}$ to $10^1$ $M^{-1} \cdot s^{-1}$) in spite of the steric stabilization of the particles, and are comparable to those of bimolecular reactions, (thereby providing a new avenue for the development of a form of chemistry where the reactants are colloids rather than molecules), (ii) elucidate the rate limiting steps of crystal-crystal coalescence leading us to propose design rules to control it, and (iii) demonstrate a simple, two-parameter model that predicts quantitatively this process and its dependence on the ligands.

Lastly, we use Brownian dynamics simulations to show how crowding effects and the relatively large size of the particles compared to their mean free path explain these remarkably large rates of coalescence and, at the same time, the puzzlingly low values of activation energy previously reported for oriented attachment processes.

# MAIN TEXT

Aggregation of crystals is usually considered an undesirable irreversible process occurring during the growth of ensembles of crystals that compromises its predictability and the control over their size and shape. Its most studied form – agglomeration – leads to larger polycrystals of usually irregular and poorly predictable size and shape[1]. Recently, another type of aggregation – oriented attachment – was reported[2,3] to occur in a crystallographically oriented fashion to form instead a larger single crystal whose shape is dictated by the orientation of attachment and the shape of the original building blocks (e.g., wires obtained from spherical particles)[4].

Soft matter (e.g., liquids, micelles, polymers), on the other hand, displays a third type of aggregation – coalescence – whereby two objects fuse and reconstruct upon successful collisions (activation energy $E_a = 10^1$-$10^2$ kJ/mol at room temperature for emulsions)[5,6] driven by the minimization of interfacial energy.

The result of coalescence is a larger object whose shape gives instead little indication of being a product of aggregation. This is why, in spite of the remarkable effort by a number of colleagues[7-13], the coalescence of crystals has remained relatively elusive when compared to that of soft matter phases. Recent work in liquid-cell transmission electron microscopes showed that some degree of reconstruction upon aggregation can indeed occur in nanocrystals[14-18], but questions remained as to whether this process does indeed occur during their synthesis in bulk solutions. The discovery and mechanistic understanding of this process in bulk solutions is unfortunately complicated by the concurrence of several mechanisms of growth (e.g., secondary nucleation, addition, aggregation, Ostwald ripening),[8,12,19-22] and the difficulty or impossibility of accurately measuring both particle sizes, shapes, and concentrations in time[7,23-26].

To overcome this challenge, we developed a model reaction system – the reaction between oleylammonium hydrosulfide (OLAHS) and $PbCl_2$ in oleylamine (OLA) to form PbS nanocrystals[27] – to investigate whether crystal-crystal coalescence occurs and in what manner. We conducted reactions by injecting OLAHS into $PbCl_2$/OLA slurries at five different temperatures (80, 100, 120, 140, and 160 °C), and collected the products at sixteen reaction times (1, 2, 4, 6, 8, 10, 12, 14, 16, 20, 24, 30, 36, 42, 50, and 60 minutes). After injection in the $PbCl_2$ slurry, the $H_2S$ liberated by dissociation of OLAHS reacts quickly to form PbS, while unreacted $H_2S$ leaves the system as a gas[27]. The lack of free sulfur precursor creates conditions of minimal supersaturation which exclude classical growth processes (i.e., supersaturation-driven addition of molecules or ions to the crystal facets). As a control experiment, we synthesized PbS nanocrystals in conditions of supersaturation-driven addition-dominated growth (cf. black triangle in Figure 1 a to d and Supporting Discussion)[28].

The reaction yield as a function of time and temperature with OLAHS (cf. Figure 1a) was approximately constant (~73 ± 15%) throughout the reaction and independent of temperature (the spread in the data is expected and is consistent with the ±20% uncertainty of the extinction coefficient)[28].

The concentration of particles as a function of time and temperature is shown in Figure 1b. The very high values observed at the end of the nucleation phase (up to ~$3.5 \cdot 10^{23}$ nanocrystals·m$^{-3}$ or ~0.5 mM) decreased rapidly with time. The number-averaged volumes of the particles as a function of time and temperature (cf. Figure 1c) shows a saturation curve with rates increasing with temperatures. Lastly, the polydispersity (cf. Figure 1d) remained remarkably constant throughout the reaction at a relatively low value of 8% regardless of the reaction temperature. Structural characterization of the products by TEM (cf. Figure 1e) and XRD (cf. Figure 1f) showed that the particles were monocrystalline, approximately spherical, and phase pure.

Aggregation does not make strong predictions on polydispersity (e.g., polydispersity can remain small if the activation energy of the process is dependent on the size of the particles), so that the relatively small polydispersity does not disprove aggregation-driven growth. On the other hand, Ostwald ripening does imply a growth of polydispersity over time that is not observed here. Lastly, the spherical shape of the particles disproves that agglomeration or oriented attachment is a dominant mechanism of growth in these conditions. Therefore, in summary, the data in Figure 1 disprove that the growth of these particles is dominated by agglomeration or oriented attachment, is inconsistent with an Ostwald-ripening process, but does not disprove coalescence. These conclusions are supported by the comparison of the analogous data from the control experiment (cf. black triangle in Figure 1 a to d and Supporting Discussion).

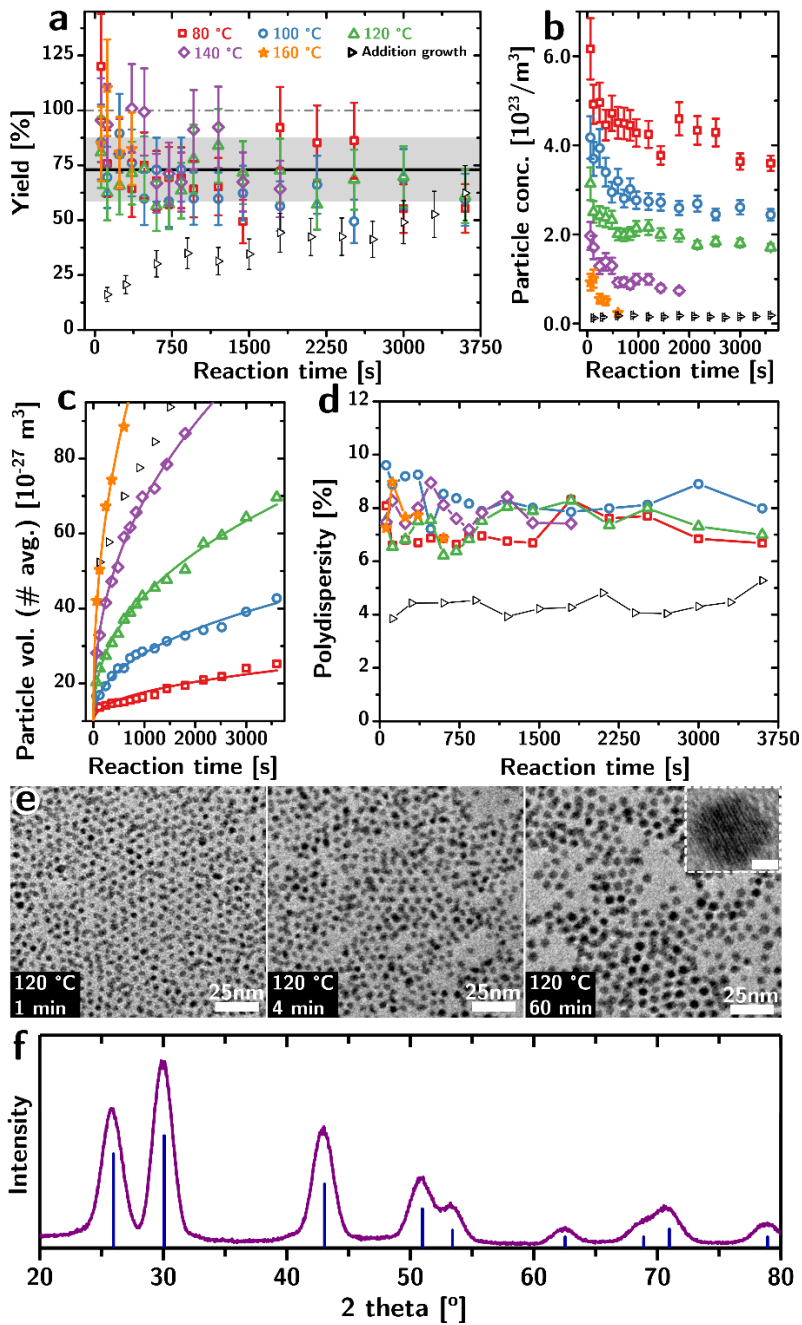

**Figure 1. Growth kinetics of PbS nanocrystals in the absence of supersaturation.** (a) Reaction yield (black line: mean, grey band: one standard deviation about mean), (b) concentration of particles (data is offset for clarity to show the decrease over time), (c) number-averaged particle volumes (globally fitted with power law), and (d) polydispersity as a function of time (abscissa), and temperature. Lines are added for clarity. For comparison, the growth kinetics of PbS nanocrystals from the control experiment (i.e. growth dominated by classical supersaturation-driven addition) is also presented in (a) to (d). (e) Representative TEM micrographs of PbS nanocrystals obtained after 1, 4, and 60 min of growth at 120 °C. Scale bar of the inset: 2 nm. (f) XRD pattern of the nanocrystals with overlaid peak positions for PbS (galena, PDF#00-005-0592). Legend for panel b-d is the same as in panel a.

Additional crucial information is revealed by the analysis of the optical absorption spectra. Figure 2a shows the evolution of the background-subtracted absorption peak associated with the 1S1S excitonic transition of PbS during growth at 100 °C in the absence of supersaturation. By comparison, Figure 2b shows the same data from the control experiment. While the control experiment shows a gradual shift to lower energies of a single Gaussian peak (a Kolmogorov-Smirnov test[29] did not reject the normal distribution), the same absorption peak in the absence of supersaturation (Figure 2a) is distinctly non-Gaussian. In fact, its features and shoulders are present at all reaction temperatures (cf. Figure 2c and Figure S1) and suggest the convolution of multiple sub-peaks, i.e., the presence of distinct populations of particles different in mean size.

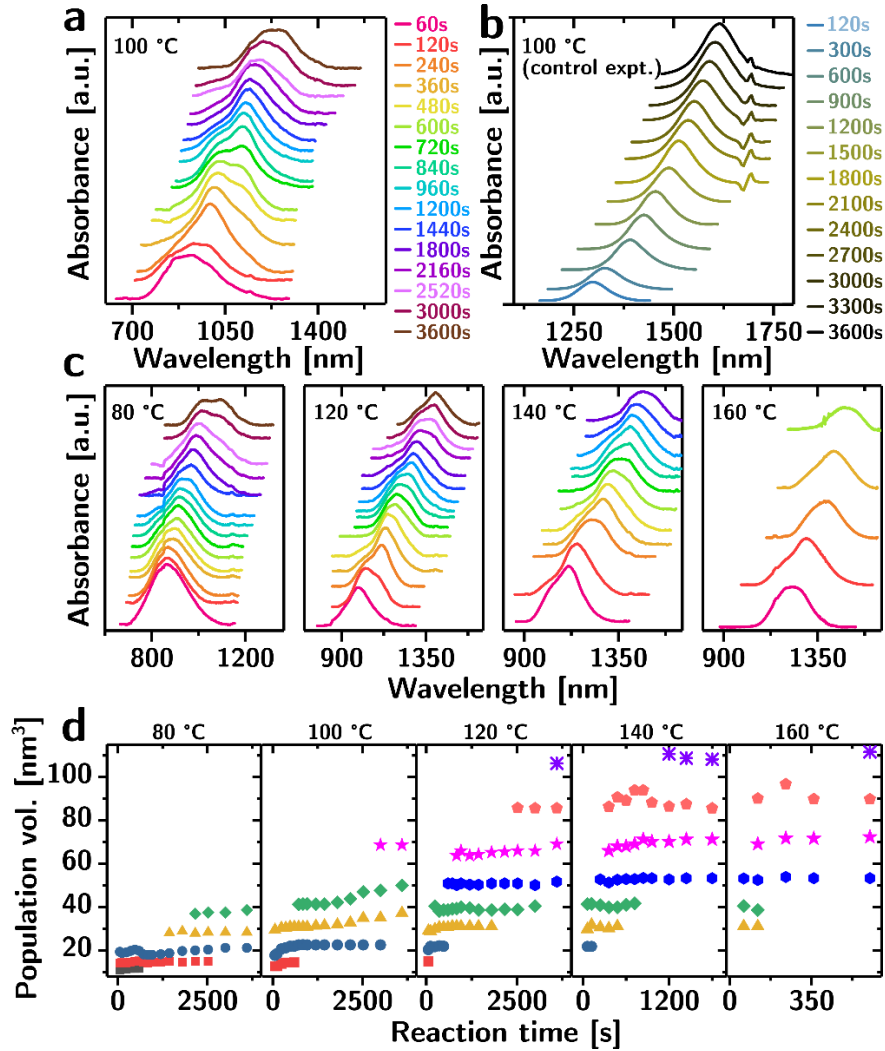

**Figure 2. Optical characterization of PbS nanocrystals grown in the absence of supersaturation.** (a-b) Comparison of the evolution during growth at 120 °C of the absorption peak from the 1S1S transition (background subtracted) in the absence (a) and (b) presence of supersaturation. (c) As panel (a) but for reactions at 80 °C, 120 °C, 140 °C, 160 °C. (d) Particle volumes of the individual populations as a function of reaction time (abscissa) and temperature (80 °C, 100 °C, 120 °C, 140 °C, 160 °C from left to right).

The data in Figure 1 disproved agglomeration and oriented attachment as driving processes in this reaction, but did not disprove coalescence or Ostwald ripening. For the growth to be caused purely by coalescence, the populations of particles generated by it should change in concentration, but not in mean size. The deconvolution of the 1S1S peaks into individual, Gaussian sub-peaks yielded, for each spectrum, a minimum set of two to four sub-peaks ($R^2>0.99$). Since the size dependence of the 1S1S transition energy and oscillator strength is known[28], each sub-peak provided us with the number-averaged volume and concentration of each detectable population.

The plot in Figure 2d is composed of 5 panels, one for each reaction temperature, from 80 °C (left) to 160 °C (right). The abscissa indicates the reaction time while the ordinate indicates the number-averaged volumes of the individual populations. The volumes of the individual populations are largely constant in time, and are strikingly conserved across reaction temperatures, which is consistent with coalescence and disproves Ostwald ripening as a dominant growth mechanism in these conditions[9] (cf. Supporting Discussion). TEM analysis cannot provide sufficient resolution and/or statistical power to identify the individual populations, but a statistical analysis of the spectra-derived particle volumes shows remarkable convergence on a value for the volume of the "monomer" particles between 9.75 and 9.87 nm$^3$ (cf. Supporting Discussion).

A growth kinetic model based on population balance equations[2,30,31] allowed us to determine the rate limiting steps of crystal-crystal coalescence. If we consider a maximum of $p$ populations (for computational convenience), then the coalescence process is described by

$$\frac{dN_n}{dt} = \begin{cases} -\sum_{m=2}^{p-1} k_{1,m} N_1 N_m, & n = 1 \\ \sum_{m=1}^{n-1} \frac{1}{2} k_{n-m,m} N_{n-m} N_m - \sum_{m=1}^{p-n} k_{n,m} N_n N_m, & 1 < n < p \\ \sum_{m=1}^{p-1} \frac{1}{2} k_{p-m,m} N_{p-m} N_m, & n = p \end{cases} \quad \text{Eq. 1}$$

where $N_i$ is the number of particles in population $i$. The kernels $k_{n,m} = Q_{n,m} e^{-\frac{E_{a,nm}}{k_B T}}$ describe the physics of the process: $Q_{nm}$ quantifies the collision frequencies, while the activation energies $E_{a,nm}$ describe the mechanism of coalescence.

Our approach was to identify the simplest coalescence model that described the data. Therefore we formulated a number of models (see Supporting Discussion) and found that the model that best described the observations was also among the simplest, with only two floating parameters, just like the Ostwald equation.

The model describes coalescence as being limited by two processes: ligand shell penetration and reconstruction (cf. Figure 3a). The first term of the activation energy

$$E_{a,nm} = \frac{C}{\left(\frac{1}{r_n+l_n}+\frac{1}{r_m+l_m}\right)} + D, \quad \text{Eq. 2}$$

can be interpreted to quantify the net energy cost of penetrating the ligand shell since it is proportional to their contact area[32]. The second term is meant to quantify the net energy cost of coalescing the particles, after ligand shell penetration has occurred. $C$ and $D$ are fitting parameters that quantify the relative importance of the two processes and are independent of particle size, time, or temperature, while $l_n$ and $l_m$ are the thicknesses of the ligand shells for the respective $n$ and $m$ populations (see Supporting Discussion for the modeling of the dependence of ligand length on particle size).

In spite of its significant (but fairly well understood) limitations,[31] we chose to use Smoluchowski's model[33] to describe the collision frequencies as

$$Q_{n,m} = \frac{4\pi(R_n+R_m)(D_n^0+D_m^0)}{W} \qquad \text{Eq. 3}$$

where $W$ is the correction factor used to account for the effect of interactions[34] (see Supporting Information), $R=r+l$, and $D_n^0$ is the diffusivity of the colloids calculated according to the Stokes-Einstein equation.

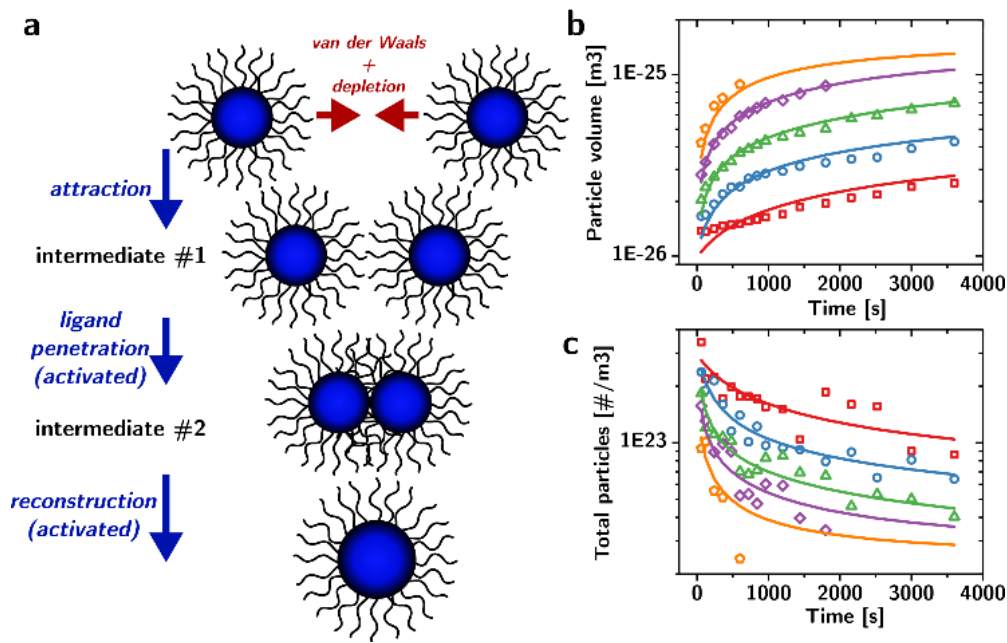

**Figure 3. Kinetic model to describe the coalescence process.** (a) Sketch of the two step process of coalescence involving ligand-shell penetration and reconstruction. (b-c) Comparison between experimental kinetic data (open scatters, red: 80 °C, blue: 100 °C, green: 120 °C, purple: 140 °C, orange: 160 °C) and the proposed kinetic model (lines) for both number-averaged particle volumes (panel b) and particle concentrations (panel c)

As shown in Figure 3b-c, our model correctly describes the temporal evolution of the particle volumes and concentrations at all temperatures characterized, using the same two values of the parameters C and D. The average ratio between the two terms of the activation energy is 0.86, indicating that the two processes are similarly important in determining the rate of coalescence. The average activation energy is $1.12 \cdot 10^{-19}$ J, or 67.65 kJ·mol$^{-1}$ which is consistent with prior reports in other material systems[23,35-38]. While this value yields rate constants that correctly describe the kinetics of coalescence, it is hard to believe that they correspond to the real activation energy of a process that makes and breaks dozens of bonds (for context, it is five times smaller than the dissociation energy of the Pb-S bond[39]). Further indication that the model does not completely capture the reality of the process lies in the fact that it overestimated polydispersities by 28% or more.

We hypothesized that the overestimation of polydispersity, and the unrealistic values of activation energy stemmed from an incorrect description of the collision frequencies by Smoluchowski's kernel (Equation 8). Our reaction environment was relatively crowded (particle volume fraction = ~3%; the average distance between the surfaces of the ligand shells ranged between 12 and 25 nm) and coalescence is thermally activated and reaction-limited. Smoluchowski's kernel does not account for crowding effects[31]. Work performed over the past few decades to correct Smoluchowski's kernel to account for crowded systems usually focused on diffusion-limited conditions and agglomeration[40,41].

Therefore, we conducted Brownian dynamics simulations to test two specific hypotheses. (i) Crowding suppresses the rate of collisions between members of rare populations by limiting the number of diffusion trajectories that can bring them into contact. (ii) Upon collision, the time spent in contact by two particles depends on their size. A depiction of these two mechanisms, which we call here respectively "traffic" and "encumbrance", is shown in Figure 4.

Figure 4a shows four diffusion trajectories that could bring the particle on the left in collision with the particle on the right. The scenario shown on the left is the one modeled by Smoluchowski's equation, i.e., the trajectories are unimpeded. The scenario on the right is closer to our experimental system: the collision trajectories are impeded by members of other, more numerous populations. Since the average rates of collisions are a function of the number of possible collision trajectories[42], the net effect should be a suppression of the rate of collisions between rarer (i.e., larger) populations and therefore a slower increase in polydispersity.

Figure 4b shows instead two particles that, at first in contact, conduct three diffusion steps representing the mean free path. On the left, the particle size is commensurate to their diffusion mean free path,[43] (as in the case for small molecules). The trajectory causes the two particles to detach and increase their mutual distance. On the right, the particles are instead much larger than the mean free path: the particles detach only to collide again, i.e., the same diffusion trajectories lead to different numbers of collisions depending on the size of the particle[44].

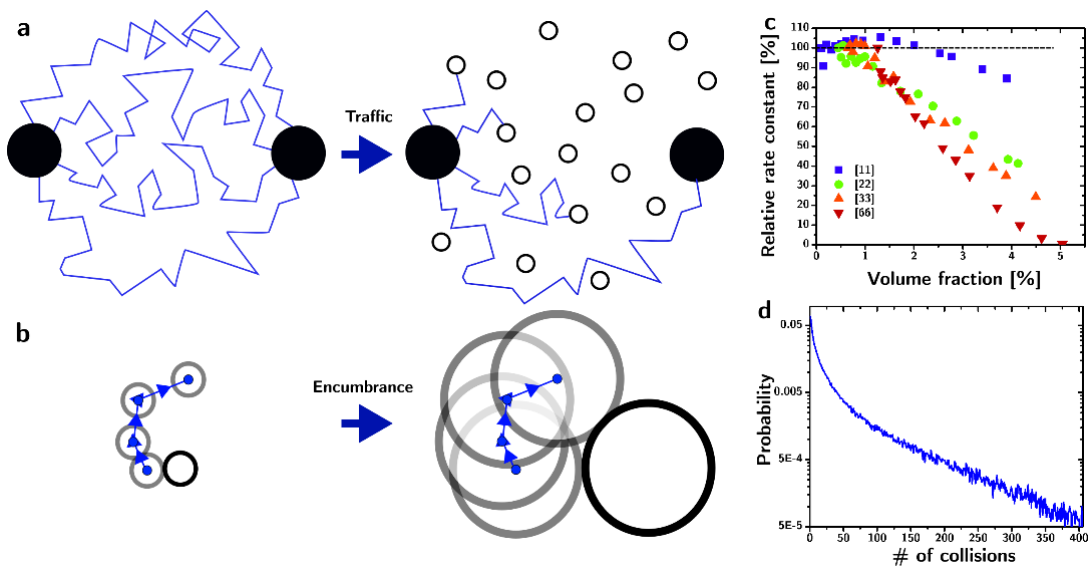

**Figure 4. Traffic and encumbrance effects on collision frequencies.** (a, left) Sketch of 4 possible diffusion trajectories (blue lines) that could cause a collision between the two black particles from the same population in the absence of particles from different populations. (a, right) Sketch of the same two particles and the same four trajectories in the presence of particles from a different population (white circles). (b, left) Sketch of a three-step trajectory bringing one particle originally in contact with another to separate. (b, right) Sketch of the same exact trajectory shown in panel c, but applied to larger particles, showing how the trajectory now causes the particles to recollide at the third step. (c) Collision rate of a fixed concentration of monomers (blue squares),

dimers (green circle), trimers (up triangle), hexamers (down triangle), as a function of the volume fraction of particles in the system (modified by adding additional monomers). (d) Probability distribution of the number of collisions observed between 2 particles initially in contact over the span of 500 diffusion steps.

To verify these hypotheses we conducted Brownian dynamics simulations[45]. In our first simulation the volume was populated with either 10 monomers, dimers, trimers, or hexamers and the volume fraction of the dispersion was then increased (the range of volume fractions was similar to the experimental one, ~3%) by adding monomers as crowders. The simulation consisted of $2\cdot10^6$ time steps, 1 ns each (2 ms of real time). Figure 4c shows the rate constants of monomer-monomer ([11]), dimer-dimer ([22]), trimer-trimer ([33]), and hexamer-hexamer ([66]) collisions as a function of the total particle volume fraction (normalized against the value obtained in the absence of fillers). The [11] collision rates are largely unaffected by the presence of the fillers (as expected, since they are also monomers). The [22], [33], and [66] collision rate constants instead drop rapidly with an increase in the total particle volume fraction. The larger the colliders, the faster their collision rate drops with an increase in the concentration of crowders.

To simulate the encumbrance hypothesis we instead started the simulation with two monomers in contact with each other. They were then allowed to diffuse (time step=$1\cdot10^{-9}$ s, 500 steps). The duration of the simulation in real time was chosen so that the RMS displacement of the particles would be <10 nm. The probability distribution of the number of collisions (cf. Figure 4d) is consistent with the hypothesis: most simulations result in a number of collisions much greater than 1. This result is fully consistent with the individual nanoparticle growth trajectories observed in liquid cell TEM[14]. Unfortunately, the distribution does not appear to be easily describable analytically, therefore compromising the accurate estimation of an average collision rate.

Since quantitatively predicting a priori the concentration and size of nanocrystals during a reaction is still a monumental challenge, we decided to test the predictive ability of our model. Therefore we conducted three separate reactions using three different ligands: octadecylamine (C18-$NH_2$), hexadecylamine (C16-$NH_2$), and tetradecylamine (C14-$NH_2$). All other parameters were kept the same as in the reactions with oleylamine (T= 120 °C, 16 time points). The number-averaged particle volumes and particle concentrations are shown by the scatters in Figure 5.

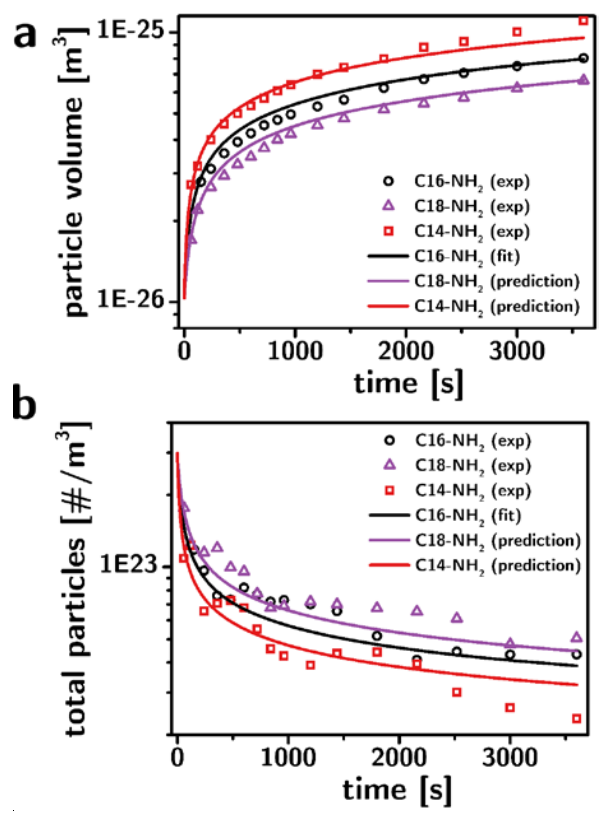

**Figure 5. Prediction of growth kinetics of PbS nanocrystals capped with n-alkylamines of different lengths at 120 °C.** (a) Number-averaged particle volumes as a function of time. (b) Particle concentrations as a function of time. Color coding indicates the amine ligand (C18-NH$_2$=red, C16-NH$_2$=black, C14-NH$_2$=purple). The scatters indicate the experimental data. The lines indicate results from the model (for C16-NH$_2$ the curve is a best fit, while for other ligands it is a prediction).

To predict the growth kinetics we first determined the values of the key variables, i.e., C and D. Since D is independent of particle size and temperature, we hypothesized that its value is associated with the inorganic phase. Therefore we used the optimal value found in the reaction with OLA (see Supporting Discussion). We used instead the data from the experiment with C16-NH$_2$ to find the value of the C parameter (3.09·10$^{-11}$ m$^{-1}$, black curves in Figure 5). Armed with the values of C and D, we used our model to predict the growth kinetics for the C18-NH$_2$ and C14-NH$_2$ reactions. The results are shown as the red and purple curves in Figure 5, which show excellent agreement with both the number-averaged volumes ($R^2 = 0.97$ for C18-NH$_2$ and $R^2 = 0.95$ for C14-NH$_2$) and concentrations ($R^2 = 0.86$ for C18-NH$_2$ and $R^2 = 0.85$ for C14-NH$_2$). The value of C for the saturated amine ligands (3.09·10$^{-11}$) is significantly larger than the value obtained from oleylamine (2.74·10$^{-11}$), indicating that saturated ligands provide a ~80% reduction in the rate of aggregation, when compared to mono-unsaturated ones of equal length.

### Author Contributions

LC designed the study. BY conducted sample characterizations. LC and BY jointly conducted the syntheses, developed the simulations, analyzed the data, and wrote the paper.

### Funding Sources

This work was supported by MSR-Intel program of Semiconductor Research Corporation under Award No. 2015-IN-2582.

**Acknowledgment**

We are grateful for generous support of this work by MSR-Intel program of Semiconductor Research Corporation under Award No. 2015-IN-2582. The authors thank Matthew G. Panthani for providing access to a UV–Vis–NIR spectrometer. We thank Alberto Passalacqua for insightful input.

# LIQUID-LIKE GROWTH OF COLLOIDAL NANOCRYSTALS BY COALESCENCE

BIN YUAN[1,2], LUDOVICO CADEMARTIRI[1,2,3*]

[1] *Department of Materials Science & Engineering, Iowa State University of Science and Technology, 2220 Hoover Hall, Ames, IA, 50011*

[2] *Department of Chemical & Biological Engineering, Iowa State University of Science and Technology, Sweeney Hall, Ames, IA, 50011*

[3] *Ames Laboratory, U.S. Department of Energy, Ames, IA, 50011*

[*] *Author to whom correspondence should be addressed: lcademar@iastate.edu*

# SUPPLEMENTARY INFORMATION

## MATERIALS AND METHODS

**Chemicals:**

Mesitylene (Aldrich, 98%), oleylamine (Acros Organics, approximate C18- content 80-90%), oleylamine (OLA, Aldrich, technical grade 70%, for large-scale (above 200 mL) synthesis), octadecylamine (C18-$NH_2$, Aldrich, technical grade 90%), hexadecylamine (C16-$NH_2$, Aldrich, technical grade 90%), tetradecylamine (C14-$NH_2$, TCI, >96.0%), hydrogen sulfide ($H_2S$, Matheson, lecture bottle, product grade), chloroform (Fisher Chemical, certified ACS), lead chloride ($PbCl_2$, Aldrich, 98%), acetone (Fisher Scientific, HPLC grade), toluene (Fisher Scientific, HPLC grade), Sulfur ($S_8$, Aldrich, 99.98% trace metal basis).

**Synthesis of PbS Nanoparticles with S8 and OLA (control experiment)**

The synthesis was carried out by following previously reported procedures[1,2]. A standard Schlenk line setup was used for all the syntheses. The reaction flask was connected to a condenser which was connected to a port of the Schlenk line.

Synthesis at 100 °C using oleylamine (OLA) as the capping ligand and $S_8$/OLA as the sulfur precursor: 8.4512 gram of $PbCl_2$ was added into a 3-neck round-bottom flask containing 20 mL OLA under stirring. Air in the reaction system was then removed using the Schlenk line setup (three cycles of vacuum/Ar). Then the flask was kept under vacuum and heated to 100 °C and maintained at 100 °C for about 10 min. Subsequently, it was kept under Ar and heated up to 160 °C and maintained there for about 3 hrs. After that, the mixture was cooled to 120 °C and 2 mL $S_8$/OLA precursor was quickly injected. The reaction temperature was lowered to 100 °C for the growth of the nanoparticles (the heating mantle was lowered right after the injection to help lower the temperature to 100 °C). Aliquots of the reaction mixture were taken at different time intervals and quenched by dilution with toluene. Purification of the nanoparticles was carried out by following similar procedures reported[1] using acetone as the non-solvent for nanoparticle precipitation.

**Synthesis of PbS Nanoparticles with OLAHS (kinetic study)**

Synthesis at different temperatures (80, 100, 120, 140, and 160 °C) with OLA as the capping ligand and oleylammonium hydrosulfide (OLAHS) as the sulfur precursor: 84.5120 gram of $PbCl_2$ was added into a 3-neck round-bottom flask containing 200 mL OLA under stirring. Air in the reaction system was then removed using the Schlenk line setup (three cycles of vacuum/Ar). Then the flask was kept under vacuum and heated to 100 °C and maintained at 100 °C for about 10 min. Subsequently, it was kept under Ar and heated up to 160 °C and maintained there for about 3 hrs. After that, the mixture was cooled to the corresponding reaction temperature (80, 100, 120, 140, or 160 °C). For the reaction at 80 °C, 40 mL of mesitylene was added to facilitate mixing. For the reaction at other temperatures, no mesitylene was added. When the temperature stabilized, 20 mL of freshly made OLAHS in mesitylene was quickly injected. After the injection, the temperature was maintained at the corresponding reaction temperature for the growth of the nanoparticles. Aliquots of the reaction mixture were taken at different time intervals and quenched by dilution with toluene. Purification of the nanoparticles was carried out by following similar procedures reported[1] using acetone as the non-solvent for nanoparticle precipitation.

**Synthesis of PbS Nanoparticles with n-alkylamines (prediction study)**

Synthesis at 120 °C using n-alkylamines (i.e. tetradecylamine, hexadecylamine, or octadecylamine) as the capping ligand and OLAHS as the sulfur precursor: 10.5640 gram of $PbCl_2$ was added into a 3-neck round-bottom flask containing 22.1624 gram n-alkylamines [the density of the n-alkylamines used here was determined to be ~0.7528 g/mL at 90~100 °C]. Air in the reaction system was then removed using the Schlenk line setup (three cycles of vacuum/Ar). Then the flask was kept under vacuum and heated to 100 °C and maintained at 100 °C for about 10 min. Subsequently, it was kept under Ar and heated up to 160 °C and maintained there for about 3 hrs. After that, the mixture was cooled to 120 °C and 10 mL mesitylene was injected. [mesitylene was added to facilitate mixing and sample collection]. When the temperature of the mixture stabilized at 120 °C, 2.5 mL of freshly made OLAHS in mesitylene was quickly injected. The reaction temperature was maintained at 120 °C for the growth of the nanoparticles. Aliquots of the reaction mixture were taken at different time intervals and quenched by dilution with chloroform. Processing of the collected samples prior to the UV-Vis-NIR measurement: (i) The collected samples were first centrifuged at 751 g for 5 min and 3004 g for 10 min, and the supernatants were collected, (ii) the collected supernatants were filtered through a 0.2 micron syringe filter.

# CHARACTERIZATION

## X-ray Diffraction (XRD)

The powder X-ray diffraction (XRD) pattern was obtained using Siemens the D500 X-ray diffractometer. The film as prepared by drop-casting the nanoparticle dispersion on a zero diffraction plate.

## UV Visibile Near Infrared (UV-Vis-NIR ) Absorption Spectroscopy

All the UV-Vis-NIR absorption spectra were obtained using a Perkin-Elmer Lambda 750 instrument.

## Transmission electron microscopy (TEM)

Images were collected using a 2007 JEOL 2100 200 kV STEM in TEM mode at 200 kV.

# DATA PROCESSING/ANALYSIS

## Calculation of the Reaction Yield, Particle Sizes, and Concentrations

The number-averaged particle concentration and the average size of the particles were obtained for each sample by the following process. First, the average size of the particles in the dispersion was calculated by taking the energy of the 1S1S exciton (E1) transition and solving the following empirical equation[1] for the core radius $r$:

$E_1[\text{eV}] = 0.41 + 0.96 \cdot r^{-2} + 0.85 \cdot r^{-1}$.    Equation S1

The average size of the nanocrystals was then used to calculate the extinction coefficient by using the empirical equation[1]

$\varepsilon[\text{M}^{-1} \cdot \text{cm}^{-1}] = 2030790 \cdot r^{2.49}$.    Equation S2

Lastly the concentration was obtained from Lambert-Beer's equation

$A = \varepsilon \cdot c \cdot l$    Equation S3

where $c$ is the concentration of nanoparticles (in M units), $l$ is optical path length (in cm units) and $A$ is the integrated absorbance of the 1S1S exciton peak.

The reaction yield was calculated according to the method reported previously[2].

## Calculation of the Nanoparticle Polydispersity from Absorption Spectra

Absorption spectrum (background subtracted) was first converted to the nanoparticle size distribution based on the previously reported relationship[1]. Then the mean and standard deviation of the size distribution were obtained, from which the polydispersity was calculated by (standard deviation/mean)·100%.

**Determination of the Length of the n-alkylamines**

The contour length of the n-alkylamines used in this study (i.e. tetradecylamine, hexadecylamine, or octadecylamine) was determined using ChemDraw (Chem3D) under energy minimization conditions.

**Determination of the Viscosity of the n-alkylamines/mesitylene Binary Reaction Mixtures at 120 °C**

The viscosities of the binary mixtures (i.e. tetradecylamine/mesitylene, hexadecylamine/mesitylene, and octadecylamine/mesitylene) were estimated using the Grunberg and Nissan equation[3]. The viscosities of the pure n-alkylamines at 120 °C were estimated by extrapolation of the previously reported data[4]. The viscosity of the pure mesitylene at 120 °C was also estimated by extrapolation[5]. The characteristic parameter G in the Grunberg and Nissan equation for mesitylene/tetradecylamine, mesitylene/hexadecylamine, and mesitylene/octadecylamine at 120 °C and at the specific mole fraction of mesitylene in the binary mixture used in this study were estimated from a closely related mixture (benzene/alkanes) by interpolation and extrapolation[3].

**Population Balance Simulations**

We developed a computational model for the growth kinetic based on population balance equations[6-8]. Consider the simplest case of a collection of $N_1$ spherical monomers of radius $r_1$ and that these monomers coalesce in a reaction-limited process to form spherical dimers of radius $r_2 = 2^{1/3} r_1$. The process is described by coupled equations describing the creation and annihilation processes:

$$\frac{dN_2}{dt} = \frac{1}{2} k_{11} N_1^2 \qquad \text{Equation S4}$$

$$-\frac{dN_1}{dt} = k_{11} N_1^2 \qquad \text{Equation S5}$$

where $k_{11}$ is the rate constant (commonly referred to as the kernel) for the coalescence between monomers defined by the Arrhenius equation

$$k_{11} = Q_{11} e^{-\frac{E_{a,11}}{k_B T}} \qquad \text{Equation S6}$$

where $Q_{11}$ is a prefactor that quantifies the collisional rate between monomers, $E_{a,11}$ is the activation energy for the coalescence of two monomers, $k_B$ is Boltzmann's constant and T is the reaction temperature.

If we consider a coalescence process that produces a maximum of $p$ populations (for computational convenience), then the system is described by the system of $p$ differential equations

$$\frac{dN_n}{dt} = \begin{cases} -\sum_{m=2}^{p-1} k_{1,m} N_1 N_m, & n = 1 \\ \sum_{m=1}^{n-1} \frac{1}{2} k_{n-m,m} N_{n-m} N_m - \sum_{m=1}^{p-n} k_{n,m} N_n N_m, & 1 < n < p \\ \sum_{m=1}^{p-1} \frac{1}{2} k_{p-m,m} N_{p-m} N_m, & n = p \end{cases} \qquad \text{Equation S7}$$

Simulations of the growth kinetics were conducted with two complementary computational approaches. In one approach we solved numerically Equation S7. This approach allowed us to include the size dependence of activation energy, but it became computationally intensive when we tried to include the effect of interactions. The other approach used a finite difference approach by discretizing the process in time and calculated the change in the concentrations of each population at each discrete time point (3600 time points for a total real time of 1 hr). The separation between time points was logarithmically distributed to compensate for the larger concentrations and faster rates at early stages of reaction: the time intervals at the beginning of the reaction were as small as $6 \cdot 10^{-9}$ s. Differently from the continuum approach, this discrete approach allowed us to include straightforwardly the influence of all interactions (especially the depletion interaction).

Code for this simulation is provided as Supporting Material.

**Brownian Dynamics Simulation**

Briefly, the particles were simulated as hard spheres. The diffusion was modeled as a random flight. The time steps equaled 1 ns, the direction of travel was randomly determined at each time step, while the distance traveled at each step was normally distributed with zero mean and variance equal to 6Dt. The RMS displacement of the particles at each time step was in the order of 2 Å. Given the fine time resolution and the small displacements at each step, we neglected changes in the direction of motion of the particles upon collision: colliding particles were returned to their original position.

To further limit the computational burden and allow for longer simulations, the simulated volume was a cube of 60 nm in size. Periodic boundary conditions were established by using a "ghost" particle approach[9].

Our code allowed us to pick a number of particles for each population (therefore determining concentrations). Collisions were detected by calculating the mutual distances between particles at each time step and comparing them to the sum of the respective radii.

# SUPPLEMENTARY FIGURES

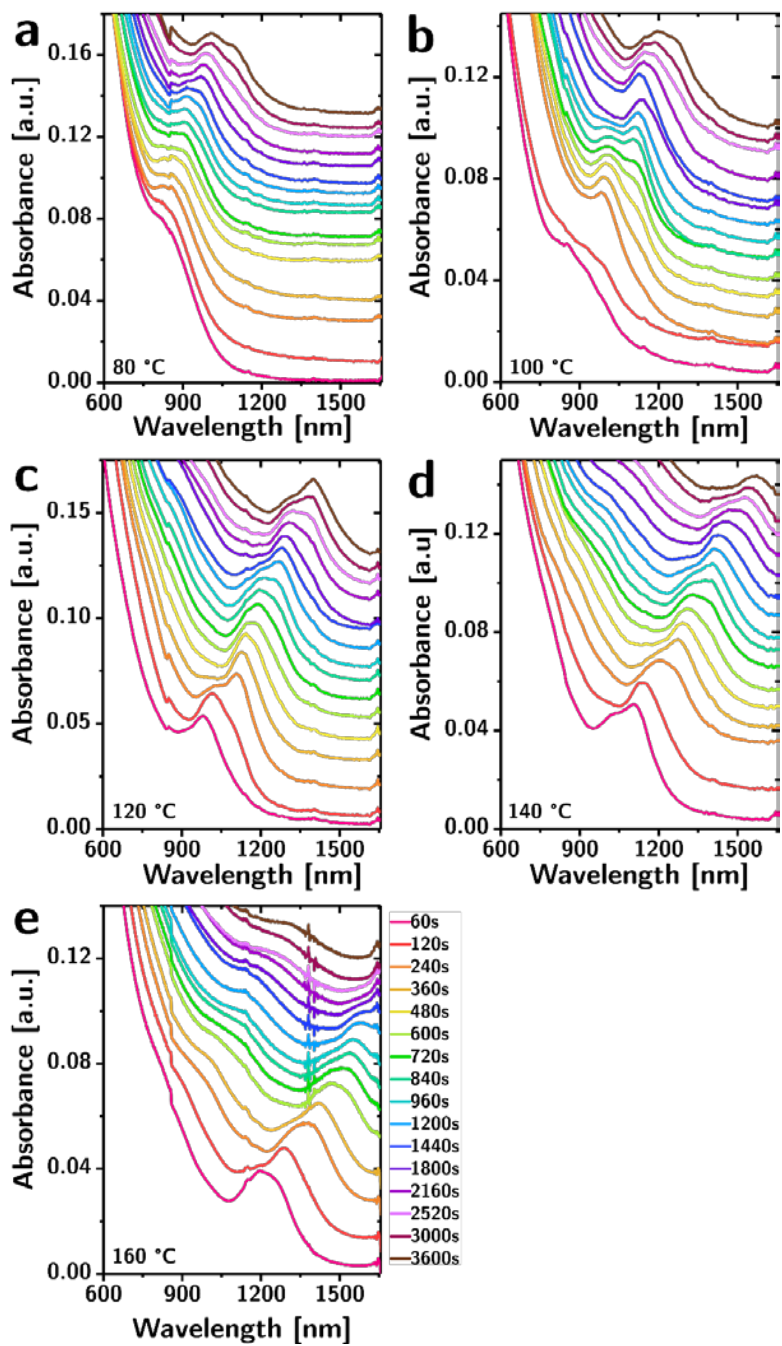

**Figure S1. UV-Vis-NIR absorption spectra of the PbS nanoparticles synthesized under different reaction temperatures with OLAHS:** (a) 80 °C, (b) 100 °C, (c) 120 °C, (d) 140 °C, and (e) 160 °C.

# SUPPLEMENTARY DISCUSSION

**Control Experiment (Growth of PbS Nanocrystals in the Presence of Supersaturation)**

Briefly, this hot-injection reaction was conducted at 100 °C by the injection of a solution of $S_8$ (1.480 mM in OLA) into a $PbCl_2$ slurry (1.417 mM in OLA). These heterogeneous reaction conditions have been shown to yield highly uniform particles through a protracted focusing (i.e., high supersaturation) phase that was caused by the gradual dissolution of the $PbCl_2$ precursor and slow reactivity of the $S_8$-OLA solution[10,11]. While the sulfur precursor is different ($S_8$ vs OLAHS), current understanding indicates that the active species in both precursors is $H_2S$[12,13], thereby making the two reactions chemically comparable.

Figure S2a shows the evolution of optical absorption spectrum of PbS nanocrystals grown in our control experiment. The peak at highest wavelength is due to the 1S1S exciton transition[14]. Other features above the background correspond to higher energy transitions[1].

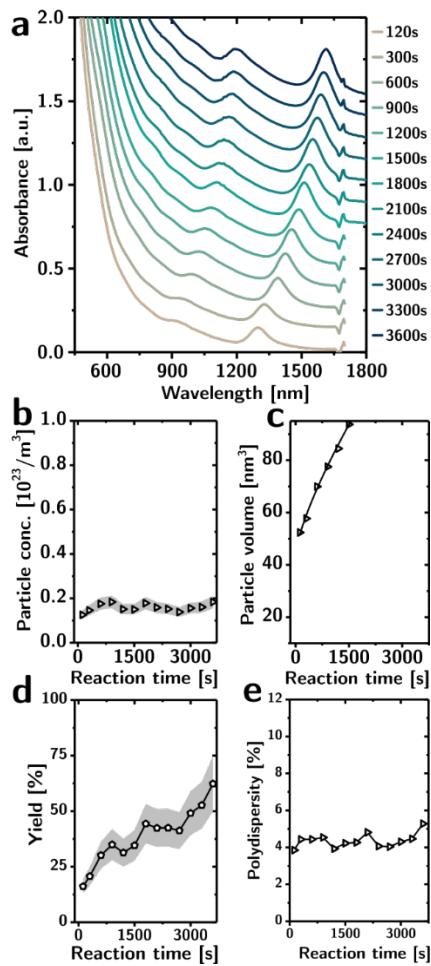

**Figure S2. Growth kinetics of PbS nanocrystals in the presence of supersaturation.** (a) Absorption spectra (offset for clarity), (b) concentration of particles, (c) number-averaged particle

volume, (d) reaction yield, and (e) polydispersity as a function of reaction time. In both panel b and d the grey area indicates the uncertainty on the mean value.

The concentration of the particles (in particles·m$^{-3}$) as a function of time (in seconds) is shown in Figure S2b. The concentration of particles does not change significantly and remains at ~$0.15·10^{23}$ particles·m$^{-3}$ (~$2.5·10^{-2}$ mM). On the other hand the number-averaged volumes of the particles (shown in Figure S2c in units of nm$^3$) increase with time as a power law that extrapolates to an initial volume of 44 nm$^3$ (diameter = 4.4 nm). Interestingly, the Ostwald ripening equation $D - D_0 = k(t - t_0)^{\frac{1}{n}}$ (where D is diameter, k is the rate constant, t is the reaction time, n is the growth exponent, and subscript 0 identifies the values of D and t at the beginning of the Ostwald ripening (i.e. the beginning of the reaction here)) fits very adequately these data ($R^2$=0.996) with a value of n (1.9±0.1) which is acceptable for Ostwald ripening[15] (between 2 and 4, depending on the rate-limiting transport process[16]).

The yield (shown in Figure S2d in percentage units) shows a steady increase from ~15% to ~60% between 1 min and 1 hr. Lastly, the size polydispersity of the nanocrystals as a function of time was estimated by converting the 1S1S absorption peak (after removal of the background absorption) into a size distribution and then obtaining the mean and standard deviation. This estimate is bound to overestimate polydispersity as it implicitly interprets the intrinsic line-width of the transition as being caused by polydispersity. The data (cf. Figure S2d) shows that the polydispersity remains constant throughout the reaction at a low value of ~4.5%.

In summary, the growth kinetics in the control experiment is consistent with classical supersaturation-driven growth. The number of particles does not change significantly during growth thereby indicating the absence of particle-generating mechanisms (e.g., secondary nucleation) or particle-depleting mechanisms (e.g., Ostwald ripening, aggregation). The steady increases in time of the particle volumes and reaction yield are consistent with a nearly constant supersaturation. Finally, the low polydispersity is consistent with the simultaneous generation of the nuclei at the beginning of the reaction and the absence of coarsening mechanisms (e.g., Ostwald ripening, aggregation, etching).

**Analysis of Subpopulation Volumes and Estimation of Monomer Volume**

If the features from the absorption spectra in Figure 2 (main manuscript) were the result of noise, the volumes identified in Figure 2d would be randomly distributed. They are not. The volumes of individual populations are shared across reaction temperatures. Furthermore, a kernel density analysis of all observed volumes (bandwidth=2 nm$^3$) show a multimodal distribution (cf. Figure S3a). If these distinct volumes were caused by a coalescence process, they would be multiples of a specific monomer volume. Therefore, we tested whether we could find a monomer volume that, if used to divide the population volumes shown as peaks in Figure S3a, would result in the minimal fractional volumes. The sum of the fractional volumes as a function of a hypothetical monomer volume (cf. Figure S3b) shows that there is a clear minimum for a value of 9.653 nm$^3$ (a radius of 1.321 nm). The plot in Figure S3c shows the modes from Figure S3a as a function of the ratio between their volumes and 9.653 nm$^3$. As shown by the droplines, the modes align qualitatively with integers, i.e., the populations identified by spectroscopy would be the dimers, trimers, tetramers, pentamers, eptamers, ennamers, and undecamers. The small discrepancies can be attributed to (i) slight differences in the size of the monomers obtained at different reaction temperatures, and (ii) the overlapping of peaks in the absorption spectrum[17]. Furthermore, it is important to note that, given that our determination of the populations is indirect and relies on a deconvolution, not all populations might have been detected, and that similarly sized populations do not necessarily have comparable concentrations[17].

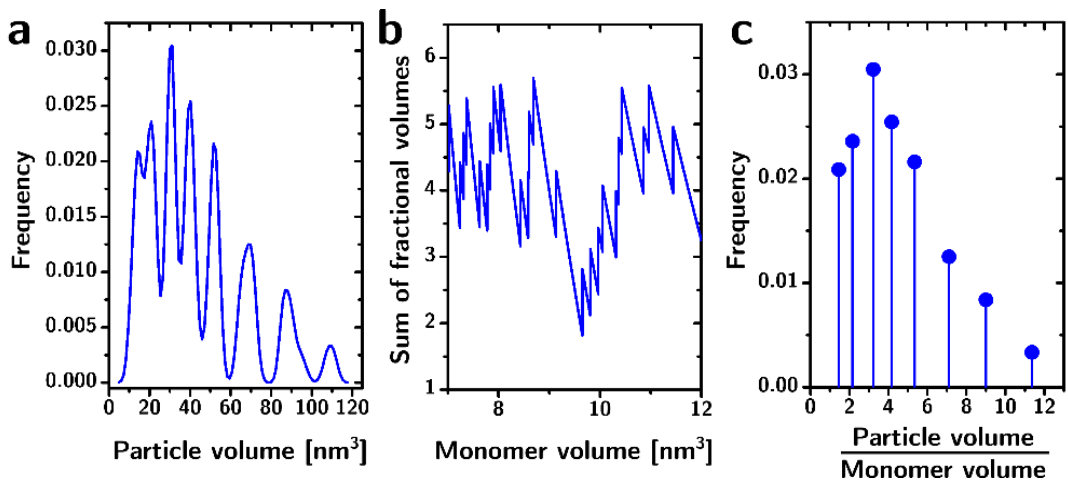

**Figure S3. Statistical analysis of the volumes of the populations.** (a) Kernel density analysis of the distribution of volumes (combined reaction times and temperatures), showing distinct peaks. (b) Sum of all fractional volumes as a function of the monomer volume, identifying an optimal monomer volumes at ~9.65 nm$^3$. (c) Peaks from panel a, rescaled as multiples of the optimal volume identified in panel b.

A separate test for the existence of a starting monomer was to extrapolate the datasets of number-averaged particle volumes vs time (cf. Figure 1c) to time = 0. To do so, we used the power law fit that is shown in Figure 1c. Remarkably, the datasets for all temperatures converge to the same starting nanocrystal volume of 9.750 ± 0.71 nm$^3$ (within 1% of the value obtained from the analysis of individual population volumes in Figure S3b).

A similar extrapolation conducted on mass-averaged volumes yields a value of 9.87 ± 0.70 nm$^3$. The ratio between the mass-averaged and number-averaged estimate for the monomer volume (i.e., the polydispersity index, PDI = 1.012 ± 0.15) suggest that the monomer is slightly polydisperse, but the uncertainty is too large to confirm it.

### *Ex-situ* TEM Evidence Cannot easily Disprove Coalescence as a Growth Mechanism

While the estimation of sizes from optical properties is well established, many colleagues in the scientific community believe that the ex situ TEM is the only trustworthy characterization that can prove the multimodality of a size-distribution. So, we did characterize a sample of those shown in Figure 1e by TEM and measured particles diameters with significant statistics (n=326). The size distribution is shown in Figure S4a as a histogram (bin size=0.19 nm) and as a kernel density function (bandwidth=0.19nm). The bin size and bandwidth were chosen to match the resolution of our TEM: one should not lightly claim features in the size distribution that are smaller than the resolution of the instrument. The resulting distribution, while non-Gaussian, does not show any obvious shoulder.

By comparison, the absorbance of the 1S1S transition of the same sample shows a clear shoulder and what is likely to be a convolution of two peaks (Figure S4b). After conversion to diameter by using Equation S1, the size distribution still shows a pronounced shoulder (Figure S4c). If one takes this optically-derived size distribution and bins the data with a resolution of 0.19 nm, one obtains a distribution (Figure S4d, solid line), which is remarkably similar to the TEM results (Figure S4d, dashed line). The difference in the average diameters between the two distributions (3.71 nm from the absorption data vs 3.90 nm from the TEM data) is fully expected. As we originally wrote in 2006 on the basis of compositional data[10], the surface of the particles is terminated by a layer of $PbCl_2$. This shell is accounted by TEM, but not by the exciton energy ($PbCl_2$ is an insulator). The difference in the radii (0.095 nm) is a third of the shortest Pb-Cl bond (0.283 nm) in the $PbCl_2$ structure, consistent with a (sub)monolayer shell.

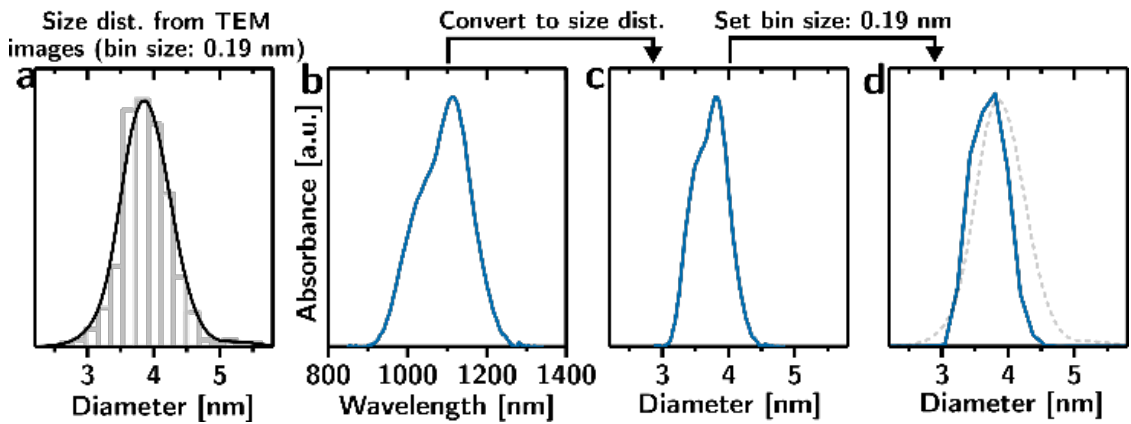

**Figure S4. Demonstration of the inadequacy of TEM-derived data in uncovering multimodal size distributions originating from coalescence.** (a) Size distribution of PbS colloidal nanocrystals obtained by coalescence (histogram and kernel density with bin size/bandwidth equal to the resolution of the microscope, 0.19 nm). (b) 1S1S absorption peak of the same sample characterized in panel a. (c) Size distribution obtained by converting the absorption data in panel b by using Equation 1. (d) Binning the distribution from panel c with a 0.19 nm bin size, and comparison with the TEM-derived distribution (dashed line).

This analysis demonstrates that a monomodal distribution obtained by a TEM analysis, even with good statistics (e.g., hundreds of particles), cannot disprove that the size distribution is bimodal/multimodal. The resolution, limitations in sample size, the relative concentrations of the concentrations between different populations, and the intrinsic limits of its size characterization (e.g., conversion of a diameter/area size information to volumes carries an error), do not allow to distinguish the different modes in the size distribution that could emerge from coalescence of particles starting from a small monomer population. For example, if the monomer has a volume of 10 $nm^3$ (i.e., diameter = 2.673 nm), the difference in radius between a pentamer and a hexamer would be 0.14 nm, and this without adding the confounding effect of polydispersity. While the physical reason why optical data can provide a higher resolution in the size distribution is that the spectrometer's resolution (~1 nm) translates into a size resolution of ~$10^{-12}$ m (homogeneous linewidth permitting), the statistical reason is that it can sample trillions of particles with each spectrum, instead of hundreds/thousands[18].

**The Experimental Results Are Inconsistent with Ostwald Ripening**

Ostwald ripening is often invoked to explain particle coarsening in conditions of low supersaturation. Our data is fundamentally inconsistent with it for the following reasons:

1. Ostwald ripening cannot cause by itself bimodal/multimodal volume distributions in fully formed crystals[19,20] (in clusters you have minima in free energy associated with specific sizes, due to a non-bulk atomic structure – in crystals there is no known mechanism why the free energy of a crystal should not change monotonically with size). The data in Figure 2 and S3 show that the dispersion is composed of polydisperse but separate populations of particles.

2. In the presence of distinct size populations, Ostwald ripening would shift the volumes of the individual populations with time. If coalescence and Ostwald ripening coexist, the size of the larger populations would gradually increase over time accompanied by the decrease in the size of the smaller populations due to Ostwald ripening[19]. We do not observe any significant shift, as shown in Figure 2d. The sharp peaks in the kernel density shown in Figure S3a are further evidence for the lack of coarsening.

3. The polydispersity is low and does not increase over time. Colloids undergoing ripening show polydispersities that are much larger than 8%: Monte Carlo simulations of nanoparticle distributions undergoing Ostwald ripening predicted a steady state polydispersity of 20%[21].

4. The Ostwald ripening equation is not diagnostic. While the particle growth kinetics shown in Figure 1c can be indeed fitted well with the Ostwald ripening equation with a physically plausible exponent *n*, the same equation also fits the growth kinetics of the control experiment, as shown in Figure S2c. The control experiment has sufficient supersaturation to cause a rapid increase in yield, and does not show a decrease in the number of particles, as expected from Ostwald ripening.

Ostwald ripening is usually invoked as the simplest explanation for particle growth in the absence of significant supersaturation. In most reported cases, the kinetic data available was not sufficient to disprove Ostwald ripening. In such cases, postulating that Ostwald ripening was the dominant growth mechanism was consistent with the principle of parsimony[22] but was inconsistent with the "burden of disproof"[23].

<u>*By comparison, our model is instead fully consistent with the observed data and relies on the same number of degrees of freedom than the Ostwald ripening model.*</u>

**Modeling the Thickness of the Ligand Shell as a Function of Particle Size**

The dependence of the thickness of a ligand shell as a function of its grafting density, Kuhn length, and contour lengths has been well described for flat surfaces[24]. At high grafting densities and in the strong stretching regime (the common scenario for ligand-capped nanocrystals), the thickness of the shell *l* can be described as

$$l = Z\left(\frac{12 b^5 \Gamma \omega}{\pi^2}\right)^{1/3} \qquad \text{Equation S8}$$

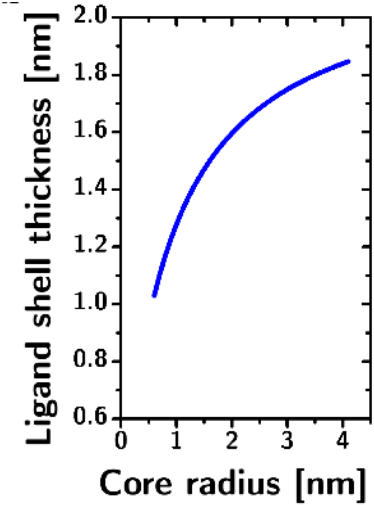

**Figure S5. Ligand shell thickness as a function of particle size.** Graph of the thickness of the ligand shell as a function of core radius.

Where $Z$ is the number of Kuhn segments in the ligand, $b$ is the length of the Kuhn segment, $\Gamma$ is the grafting density, and $\omega$ is the excluded volume parameter. Equation S8 does not consider the curvature of the surface, which can have a significant influence on the conformation of ligands[25]: for equal grafting densities, a convex surface provides the ligands with more accessible volume (and therefore, more entropy) than a flat surface. The volume available for each ligand determines its stretching. Recent reported models for ligand shells on nanocrystals focused on the dry state[26], while our experiments take place in solution, where partial solvation can significantly change the ligand shell thickness[27].

Our approach to account for curvature is to estimate the grafting density on a flat surface that would provide the ligands with the same amount of available volume that they have on surface of the nanocrystal. In brief, we calculated the grafting density on the particles by assuming 33% coverage, i.e., 33% of the exposed lead ions are coordinated by one oleylamine molecule. (Incomplete coverage is common in colloidal nanocrystals[28-30].) Then we calculated, as a function of particle size, the number of ligands per particle, and the amount of volume available to the bound ligands (i.e., the volume of a shell as thick as the contour length of oleylamine, $l_0$, divided by the number of ligands). We then calculated how many ligands would be grafted on an equivalent but flat surface if they had the same amount of volume available. On the basis of this number we could calculate the effective grafting density to be used in Equation S8. This procedure yields the following analytical function of the ligand shell thickness as a function of the particle radius

$$l(r) = Z \left( \frac{36\omega\Gamma b^5}{\pi^2} \frac{r^2}{3r^2 + 3rl_0 + l_0^2} \right)^{\frac{1}{3}} \qquad \text{Equation S9}$$

For surface-bound oleylamine in oleylamine (our reaction conditions) we used a Kuhn length of 14 Å (i.e., N=1.62) and an excluded volume parameter of 0.2[31] and obtained the data shown in Figure S5. The data is plausible: as $r$ tends to infinity, the value of $l(r)$ approaches $l_0$.

**Accounting for the Effect of Viscosity**

$D^0_n$ is the diffusivity of the colloids calculated according to the Stokes-Einstein equation (assuming spherical particle shapes) $D^0_n = k_B T/6\pi\eta R_n$, where $\eta$ is the dynamic viscosity, which depends on temperature according to a phenomenological equation reported for the similar molecule oleic acid[32]

$$\eta(T) = 3.18 \cdot 10^{-3} + 1.153 \cdot T^{-11.02} \qquad \text{Equation S10}$$

**Accounting for the Effect of Interactions on the Collision Frequencies**

Interactions can significantly modify the collision frequencies. For a given potential between the particles $E(x)$, the correction factor $W_{nm}$ [33] is equal to

$$W_{nm} = (r_n + r_m) \int_{r_n+r_m}^{\infty} exp\left[\frac{E(x)}{k_B T}\right] \frac{dx}{x^2} \qquad \text{Equation S11}$$

where $x$ is the distance between the particles. The van der Waals potential between the particles was modeled as described by Hamaker[34], as

$$E_{vdW}(x) = -\frac{H}{6}\left[\frac{2r_n r_m}{x^2-(r_n+r_m)^2} + \frac{2r_n r_m}{x^2-(r_n-r_m)^2} + ln\left(\frac{x^2-(r_n+r_m)^2}{x^2-(r_n-r_m)^2}\right)\right] \qquad \text{Equation S12}$$

where H is the Hamaker constant ($8 \cdot 10^{-20}$ J for PbS[35] and $5.2 \cdot 10^{-20}$ J for the ligands[36], across vacuum, which leads to a value of $3 \cdot 10^{-21}$ for PbS across oleylamine)[25,37].

Viscous interactions describe how friction coefficients of particles approaching each other can be quite significantly different than those of isolated particles. This effect is usually accounted for by a "diffusion ratio" $D^\infty/D_{nm}$ that multiplies the exponential term in the integral of Equation S11. The diffusion ratio was calculated, according to Spielman[38], as

$$\frac{D^\infty}{D_{nm}} = 1 + \frac{2.6 R_n R_m}{(R_n+R_m)^2}\sqrt{\frac{R_n R_m}{(R_n+R_m)(x-R_n-R_m)}} + \frac{R_n R_m}{(R_n+R_m)(x-R_n-R_m)} \qquad \text{Equation S13}$$

The Coulombic interaction between the nanocrystals was neglected since zeta potential measurements indicated the absence of charges on the particles.

Lastly, the depletion interaction between particles can be quite significant in concentrated solutions[39,40]. To describe these attractive component of the interactions we used the Asakura-Oosawa model[41]

$$E_{dep,n}(x) = \sum_{\substack{i=1 \\ i \neq n}}^{p} N_i k_B T V_{dep,nm}(x) \qquad \text{Equation S14}$$

where $V_{dep}$ is the depletion volume created when particles approach each other.

**Simulations of Growth Kinetics (Survey of Models Tested)**

To account for possible inadequacies of the corrections to the collision frequency, we included in the model a size dependent correction factor. The resulting general kernel is

$$k_{nm} = \frac{4\pi(R_n+R_m)(D_n+D_m)}{W_{nm}[A(r_n^B+r_m^B)]} \exp\frac{-\left[\frac{C}{\left(\frac{1}{R_n}+\frac{1}{R_m}\right)}+D(r_n^E+r_m^E)\right]}{k_B T} \qquad \text{Equation S15}$$

where A, B, C, D, and E are fitting parameters. To identify the minimum number of variables (i.e., degrees of freedom) necessary to describe the experimental data, we conducted simulations using versions of Equation S15 with different numbers and combinations of degrees of freedom, as described by Table SI.

The models are significantly different in the way they describe the physics of the coalescence process. In terms of collision frequencies, models #1 to #4 assume that they are adequately described by Smoluchowski's model. Model #5 and #6 instead correct them with a size-dependent power law. In terms of rate limiting mechanisms (e.g., ligand-penetration vs reconstruction), model #1 assumes that reconstruction is the slow step but is independent of size, model #2 assumes that ligand-penetration is the slow step, models #3 and #5 assume that the reconstruction step is rate limiting but is dependent on the size of the particles, while models #4 and #6 assumes that both mechanisms proceed at similar speeds.

**Table SI. Summary of the models considered for comparison with the data**

| Model # | Degrees of freedom | A: Correction to the collision frequencies | B | C: Ligand-penetration | D: Reconstruction | E |
|---|---|---|---|---|---|---|
| 1 | 1 | ✗ | ✗ | ✗ | ✓ | ✗ |
| 2 | 1 | ✗ | ✗ | ✓ | ✗ | ✗ |
| 3 | 2 | ✗ | ✗ | ✗ | ✓ | ✓ |
| 4 | 3 | ✗ | ✗ | ✓ | ✓ | ✓ |
| 5 | 4 | ✓ | ✓ | ✗ | ✓ | ✓ |
| 6 | 5 | ✓ | ✓ | ✓ | ✓ | ✓ |

The comparison between the simulated and experimental data included both the total particle concentrations and the number-averaged particle volumes (each taken as a function of time and temperature). The optimal agreement between simulation and experimental data was not obtained by a least square fitting process due to concerns that it would converge to local minima in the large and complex parameter space. We instead evaluated how variances and coefficients of determination ($R^2$) vary with combinatorial screening conditions. To limit the number of possible combinations, we assumed that B and E were integers between -3 and 3.

The best fits are shown in Figure S6 while the respective values of $R^2$, of each fitting parameter, and of the average activation energy (in units of kJ·mol$^{-1}$) are listed in Table SII. The fits shown in Figure S6 and listed in the Table include the corrections to the collision frequencies due to van der Waals, viscosity, and depletion interactions. Nonetheless those corrections made only a small difference in the fits and the optimal conditions were consistent with those found by solving the system of differential equations.

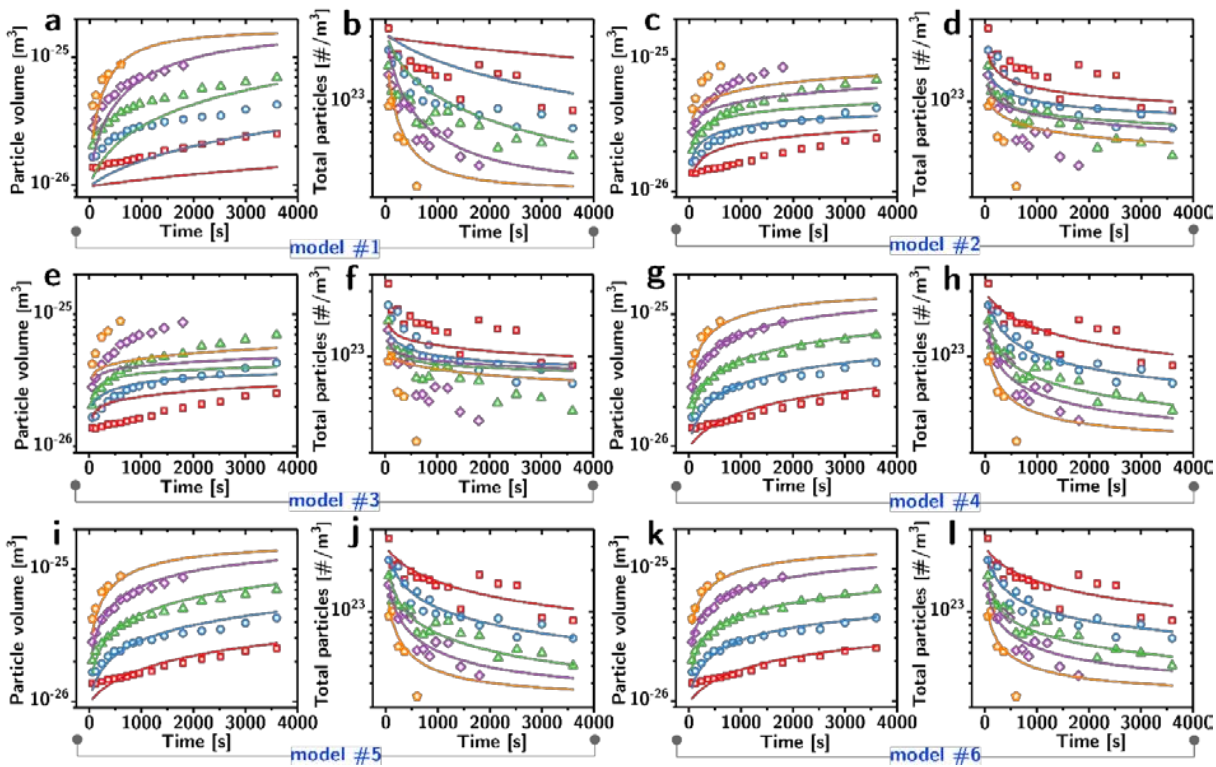

**Figure S6. Comparison of simulations with experimental data for different models.** This figure shows how the data of number-averaged particle volume and particle concentration (red: 80 °C, blue: 100 °C, green: 120 °C, purple: 140 °C, orange: 160 °C) are described by the six models described in the text. (a-b) model #1. (c-d) model #2. (e-f) model #3. (g-h) model #4. (i-j) model #5. (k-l) model #6.

**Table SII. Best fits of experimental data for the 6 models explored.**

| Model # | R² (N) | R² (V) | A | B | C | D | E | $<E_a>$ [kJ·mol⁻¹] |
|---|---|---|---|---|---|---|---|---|
| 1 | 0.21 | 0.58 | | | | 5.2E-20 | | 62.64 |
| 2 | 0.65 | 0.70 | | | 6.07E-11 | | | 79.84 |
| 3 | 0.50 | 0.44 | | | | 2.92E-11 | 1 | 96.15 |
| 4 | 0.89 | 0.99 | | | 2.74E-11 | 2.97E-20 | 0 | 67.65 |
| 5 | 0.89 | 0.97 | 3.52E+27 | 3 | | 3.99E-20 | 0 | 48.02 |
| 6 | 0.89 | 0.99 | 7.00 | 0 | 1.97E-11 | 2.64E-20 | 0 | 57.78 |

The first three models fail to describe the experimental data and won't be discussed further. Model #4 had the best values of R², and, importantly, the optimal value of the exponent E was found to be zero. Therefore, this model effectively has only two degrees of freedom. Model #5 obtained the third best values of R², but, as shown in Figure S6, it fails to accurately describe the particle volumes at long reaction times. The optimal value of E for this model was also zero, therefore reducing the number of degrees of freedom of this model to three. Model #6 obtained analogous R² values to #4. The optimal values for the two exponents (B and E) were zero therefore reducing the number of degrees of freedom for this model to three.

According to Ockham's razor[22,42], the model with the smallest number of degrees of freedom that adequately describes the experimental data should be considered the closest approximation to the truth.

Figure S7 shows the dependence of the values of $R^2$ on the effective degrees of freedom of the models, and shows how model #4 (Equation S16) appears to be the best compromise between describing accurately the data and decreasing the numbers of degrees of freedom in the model.

$$k_{nm} = \frac{4\pi(R_n+R_m)(D_n+D_m)}{W} \exp \frac{-\left[\frac{C}{\left(\frac{1}{R_n}+\frac{1}{R_m}\right)}+D\right]}{k_B T} \qquad \text{Equation S16}$$

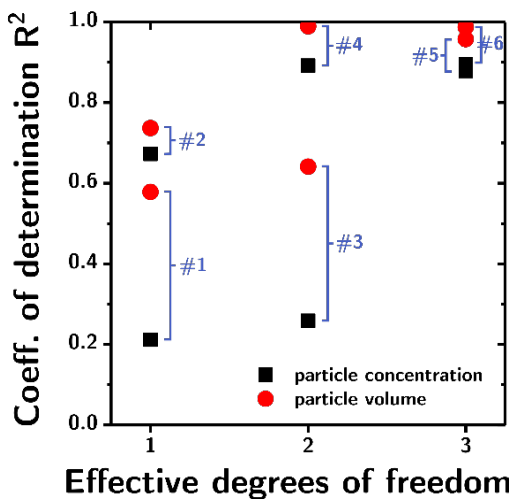

**Figure S7. Model selection.** This graph shows the best $R^2$ values obtained by each model for particle concentration (black squares) and particle volume (red circles) as a function of the effective degrees of freedom of the models.

**Conclusions**

This work reaches a number of technical, scientific, and methodological conclusions. For convenience we summarize them here along with the key pieces of evidence supporting each of them.

*Coalescence between ligand-capped nanocrystals can happen at rates comparable to bimolecular reactions.* The average rates of coalescence with oleylamine as a capping ligand ($10^{-2}$ to $10^1$ $M^{-1} \cdot s^{-1}$) were slightly lower than those observed in the end-to-end coupling of $Bi_2S_3$ colloidal nanowires ($4 \cdot 10^3$ $M^{-1} \cdot s^{-1}$), and are within the range of characteristic bimolecular reaction rates (from ~$10^{-5}$ $M^{-1} \cdot s^{-1}$ in step-growth polymerization[43] to ~$10^5$ $M^{-1} \cdot s^{-1}$ in diffusion-controlled coupling[44,45]). The lack of coalescence in the control experiment is explained by the lower concentration of particles ($2.5 \cdot 10^{-2}$ mM vs 0.5 mM, resulting in a 400-fold decrease in the rate of coalescence). These rates of coalescence suggest the tantalizing possibility that colloid-colloid "reactions" that could mimic bimolecular reactions might have been overlooked because of the relatively low concentrations of nanocrystals commonly used in nanocrystal chemistry (typically <$10^{-5}$ M, which would results in rates of aggregation ~100 slower than in our synthesis).

*The rate constants can be described quantitatively and predicted.* Our model successfully describes and predict experimental data by introducing a size-dependence to the activation energy that is proportional to contact area between ligand shells. We attribute this term of the activation energy to the process of penetrating the ligand shell. Our model does not make assumptions based on the composition of the nanoparticles, which suggests that it might be generally applicable to other materials.

*Saturated ligands appear to reduce the rate of aggregation by 80% when compared to mono-unsaturated ligands.* The model works equally well for saturated and unsaturated ligands. The difference in the activation energies between these two types of ligands indicates that saturated ligands are 80% more likely to prevent coalescence in a typical collision, possibly due to increased ligand-ligand interactions within the shell.

*Aggregation-driven growth does not need to increase polydispersity.* We have shown that aggregation-driven growth does not necessarily increase polydispersity because (i) it can result in coalescence (i.e., reconstruction of the particle into an isotropic shape), and (ii) the rates of coalescence can be strongly dependent on particle size. This realization suggests the remarkable possibility of using coalescence as a highly sustainable and scalable particle growth process requiring minimal chemical input (i.e. coalescence that takes place significantly at high concentration of particles).

*The activation energies obtained from the model are, most likely, ́effectivę́and result from an inaccurate description of the collision frequencies.* The values of the average activation energies we have obtained from the model (67.65 kJ·mol$^{-1}$) are consistent with prior literature but are not credible if taken at face value. They suggest that a process as complex as the penetration of a ligand shell and the reconstruction of an entire nanocrystal has a net energy cost that is 5 times smaller than the binding energy between the Pb and S atoms. Nonetheless, these activation energies and the resulting rates explain and predict experimental data. Our explanation is that these values of activation energy are "effective" and result from an incorrect modeling of collision rates.

*Smoluchowskís model is inadequate in describing the collisional frequencies in crowded colloidal dispersions.* We have shown how the Smoluchowski's model for collision rates neglects at least two fundamental processes that are characteristic of Brownian diffusion of colloids in crowded dispersions. We have called these two processes "encumbrance" and "traffic". The encumbrance mechanism significantly increases the time particles spends in proximity of each other between their initial collision and their separation back into the bulk solution. The mutual steric encumbrance of the particles and their large size compared to their mean free path of diffusion limit the number of trajectories that lead them to separate. The traffic mechanism instead shows that the collisions [jj] between members of a rare population j is severely suppressed by the presence of members of a common population i. Most trajectories that would lead to [jj] collisions in the absence of i particles result instead in [ij] collisions.

*It might not be possible to conduct fully ab-initio predictions of growth kinetics in ensembles of crystals in solution.* The fundamental problem of describing the kinetics of these collisions lies in the extremely long tail of the distribution of residence times of particles in contact with each other. These tails have disproportionate effect on the average kinetics of the system and could possibly be extremely sensitive to experimental parameters. In such case, the extraction of the "real" activation energies of coalescence could prove elusive.

*The Ostwald ripening equation is not diagnostic.* As we have shown in this work, the Ostwald ripening equation is sufficiently vague and flexible to allow for the fitting of growth kinetics resulting from completely different non-ripening processes (addition-based classical crystallization in the presence of supersaturation as well as aggregation-driven growth in the absence of supersaturation). We believe Ostwald ripening should only be claimed when other mechanisms have been disproven, on the basis of the principle of the "burden of disproof"[23]. In many occasions, aggregation-driven growth processes are likely to have been misinterpreted in the past as Ostwald ripening.

*Electron microscopy cannot easily disprove aggregation-driven growth.* We have shown in this work how the coalescence of small monomer particles results in differences in diameter between "neighboring" populations (e.g., hexamers vs heptamers) that are smaller than the resolution of most imaging platforms. Even if the resolution of imaging is at the single atom level, one has to account for the errors associated with the conversion of diameter/area information into volume, and the necessity of very significant sample sizes to rigorously disprove the multimodality of a distribution.

*Accurate information on the temperature-dependent viscosities of the reaction media is essential to model growth kinetics.* One often overlooked variable in studies of aggregation is viscosity, which plays a very significant role in determining the diffusivity of the particles. For studies of this kind to become increasingly quantitative it is essential to have reliable data on viscosities, especially when complex mixtures are used.

This paper shows how the exceptional optical properties of colloidal quantum dots provide unique opportunities for the study of crystal growth kinetics in ensembles of crystals, and the elucidation of complex effects (e.g., steric stabilization by ligands, crowding).